\documentclass[runningheads]{llncs}
\usepackage{graphicx}

\newcommand{\be}{\begin{equation}}
\newcommand{\ee}{\end{equation}}
\newcommand{\bse}{\begin{subequations}}
\newcommand{\ese}{\end{subequations}}
\newcommand{\bea}{\begin{eqnarray}}
\newcommand{\eea}{\end{eqnarray}}
\newcommand{\ba}{\begin{array}}
\newcommand{\ea}{\end{array}}
\begin{document}
\title{Challenges in fluid flow simulations using Exascale computing}
%
%
\author{Mahendra K. Verma\inst{1}\orcidID{0000-0002-3380-4561} }
\authorrunning{Mahendra K. Verma}
\institute{Department of Physics, Indian Institute of Technology Kanpur, Kanpur, India 208016
\email{mkv@iitk.ac.in}\\
\url{http://turbulencehub.org}} %
\maketitle              
\begin{abstract}
In this paper, I discuss  the challenges in porting hydrodynamic codes to futuristic exascale HPC systems.  In particular, we describe the computational complexities of finite difference method, pseudo-spectral method, and Fast Fourier Transform (FFT).  We show how the global data communication among the processors brings down the efficiency of pseudo-spectral codes and FFT.   It is argued that FFT scaling may saturate at 1/2 million processors.  However, finite difference and finite volume codes scale well beyond million processors, hence they are likely candidates to be tried on exascale systems.  The codes based on spectral-element and Fourier continuation, that are more accurate than finite difference, could also scale well on such systems.

\end{abstract}
\section{Introduction}

High performance computing (HPC) or supercomputing is an interdisciplinary area of research.  In addition to strong proficiency in the scientific domain and numerical algorithms, scientists and engineers working in HPC need strong programming skills, as well as good knowledge of state-of-the-art  computing hardware and software.   What makes it even more challenging is the rapidly evolving  computer hardware and software technology in a race for exascale HPC systems.   In this article, I will present the challenges faced by  computational fluid dynamists while using state-of-the-art supercomputers.   Here, we explore how some applications could possibly be scaled to exascale systems.   This  paper is based on the talk I gave in the conference ``Software Challenges to Exascale Computing (SCEC)" held in Delhi on 13-14 December 2018.


Computational fluid dynamics, CFD in short, is a major field of science and engineering with wide applications in weather predictions;   in modelling interiors and atmospheres of stars and planets, and their climate; in modelling flows in rivers, oceans, and astrophysical objects (in galaxies, blackholes); in designing and optimising automobiles and airplanes; in space technology; in petrochemical industry; in engineering appliances such as turbines, engines, etc.  Also, simulations are used for developing understanding and modelling turbulence that remains an unsolved problems till date. These CFD simulations consumes a large  fraction of computing resources in major HPC systems of the world.  Given this, it is important to design large scale CFD applications that can run on futuristic exascale machines.

The leading methods of CFD are finite difference, finite volume, finite elements, spectral or pseudo-spectral, spectral elements, vortex, etc.,  each of which have their advantages and disadvantages~\cite{Anderson:book:CFD,Ferziger:book:CFD}.  For example, a spectral method is very accurate, but it is suitable for simulating flows  only in idealised geometries such as cubes, cylinder, spheres~\cite{Boyd:book:Spectral,Canuto:book:SpectralFluid}.  In addition, its parallel version is inefficient due to \texttt{MPI\_Alltoall} communication of data. In comparison, finite difference and finite volume schemes are  less accurate, but they can simulate flows in complex geometries.  In addition, the finite difference and finite volume schemes are more efficient for parallelisation compared to spectral method.  

Simulation of complex flows, specially in turbulent regime, involves a large number of mesh points, going up to trillions.   For example, a spectral simulation on $8192^3$ grid has approximately trillion variables (velocity and pressure fields at the mesh points).  A major challenge in HPC is design CFD codes for exascale systems that will have millions of processor connected via a network of interconnects.  In this paper we will illustrate the parallelisation strategy for two schemes:  finite difference and spectral, and contrast their performance and limitations.  

We illustrate the above methods for incompressible Navier Stokes equations, which are
\bea
\frac{\partial {\bf u}}{\partial t} + {\bf u} \cdot \nabla {\bf u}  &= &
- \nabla p +\nu \nabla^2 {\bf u}, \label{eq:NS_u} \\
 \nabla \cdot \mathbf{u} & = & 0, \label{eq:div_u_eq_0ndim}
\eea
where ${\bf u}({\bf r},t)$ is the velocity field, $p({\bf r},t)$ is the pressure field,  $\nu$ is the kinematic viscosity, and $\rho$ is the  density of the fluid.  In the incompressible limit, $\rho$ is constant.  In the following two sections we will describe the parallel complexity of finite difference and spectral techniques, as well as  that of Fast Fourier Transform (FFT) that takes 70\% to 80\% of the total time in a spectral code.  We skip many algorithmic details of these methods.  For example, we do not describe how the pressure field is computed for an incompressible flow.  The reader is referred to  \cite{Anderson:book:CFD,Ferziger:book:CFD}   for  details.

Let us review some of the key CFD works.   There are several FFT libraries available at present.  Multicore-based FFTs are P3DFFT~\cite{Pekurovsky:SIAM2012}, PFFT~\cite{Pippig:CP2010}, FFTK~\cite{Chatterjee:JPDC2018}, and hybrid (MPI + OpenMP) FFT~\cite{Mininni:PC2011}.  There are several GPU-based FFTs too~\cite{Czechowski:CP2012}.  These libraries have been scaled up to several hundred thousand processors.  For examples, FFTK scales approximately up to 196608 cores of Cray XC40~\cite{Chatterjee:JPDC2018}.   Refer to Aseeri et al.~\cite{Aseeri:CP2015} for a summary of parallel scaling of the above FFT libraries and some others. 

There are many large-scales pesudo-spectral simulations too.  Here we list only a couple of them.  In 2002 itself, Yokokawa et al.~\cite{Yokokawa:CP2002} performed a turbulence simulation on $4096^3$ grid using {\em Earth Simulator}.   Chatterjee et al.~\cite{Chatterjee:JPDC2018} and Verma et al.~\cite{Verma:NJP2017} performed simulations of hydrodynamic turbulence and turbulent thermal convection on $4096^3$ grid. Yeung et al.~\cite{Yeung:PNAS2015}  performed $8192^3$ grid simulation using 262144 cores of Blue Waters, a Cray XE/XK machine.    Ishihara et al.~\cite{Ishihara:PRF2016} have a  record for a pseudo-spectral simulation with maximum grid size---$12299^3$. Another notable high-resolution spectral simulation is by Rosenberg et al.~\cite{Rosenberg:PF2015}. 

Number of numerical simulations using finite-difference and finite-volume methods are many more than that using pseudo-spectral method.    A popular finite-difference based astrophysical code is ZEUS, which is detailed in Stone and Norman~\cite{Stone:ApJ1992}.  Some other major codes in this category are by Balsara~\cite{Balsara:ApJ2004} and Samtaney et al.~\cite{Samtaney:PF2001}.   A major achievement that fetched Yang et al.~\cite{Yang:SC2016}  a Gordon Bell prize in 2016  is  a performance of numerical simulation of a weather code using 10 million cores.  We apologetically skip many other works that merits mention here.

In the next two section we will describe respectively the finite difference and pseudo-spectral schemes.

\section{Flow solvers based on finite difference scheme}
\label{sec:FD}

In finite difference scheme, the real space domain is discretized; the grid points are labelled as $(i,j,k)$, where $i, j$, and $k$ are integers.  The grid spacing is denoted by $(\Delta x, \Delta y, \Delta z)$, hence, the real space coordinates for the grid point $(i,j,k)$ is $(i \Delta x, i \Delta y, k \Delta z)$.  We assume the field variables to be represent at the grid points\footnote{This simple arrangement is called {\em collocation grid}, in contrast to more complex one called {\em staggered grid} in which the velocity fields are represented at the face centres, and pressure at the centre of the cube.  In this paper, for simplicity, we assume collocation grid.}.

\begin{figure}
\centering
\includegraphics[scale=0.5]{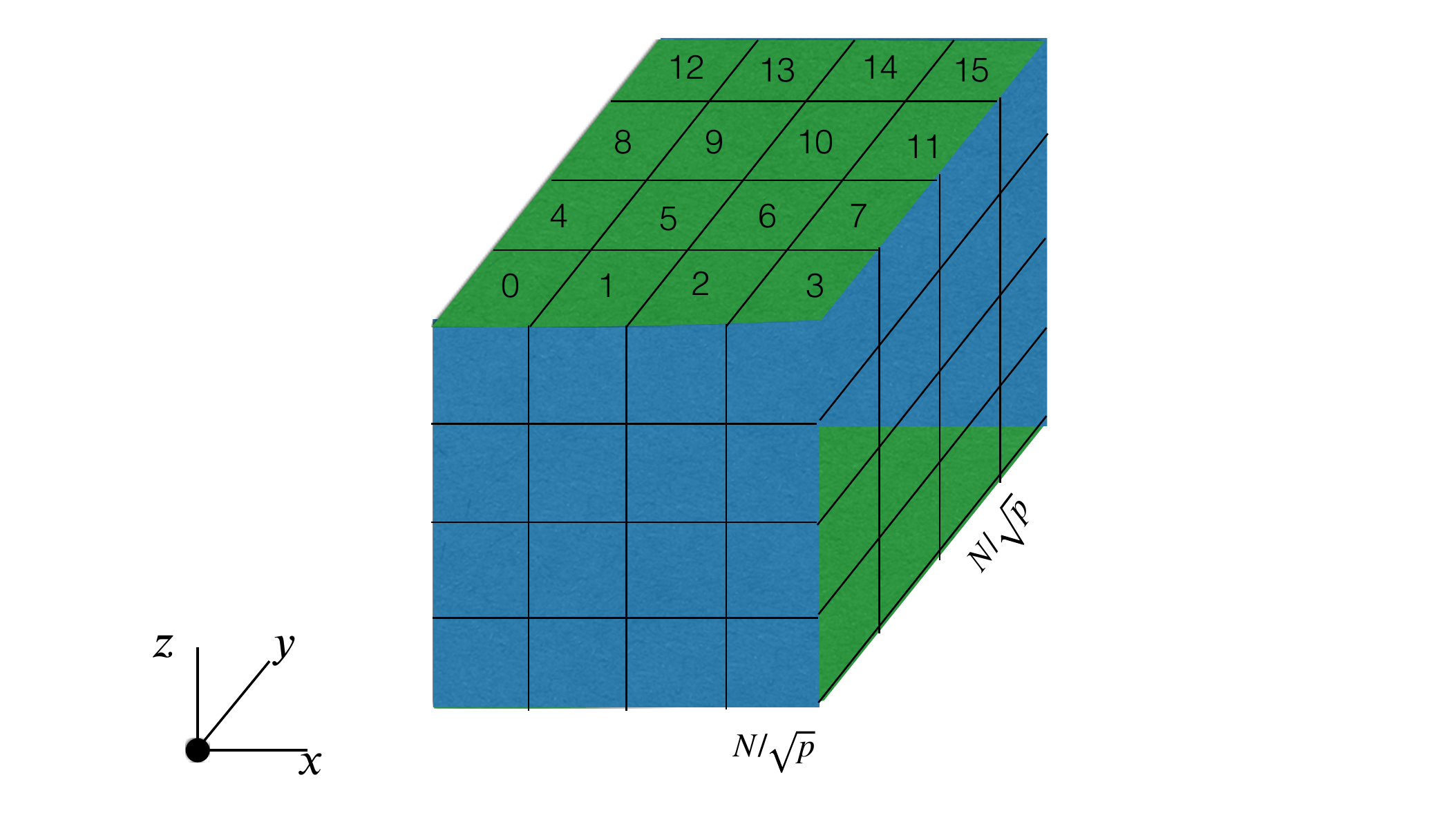}
\caption{Pencil decomposition of $N^3$ grid points among $p$ processors.  The processors, numbered as 0 to 15, are divided equally among $x$ and $y$ axis.  Each process has $N/\sqrt{p}\times N/\sqrt{p} \times N$ grid points (apart from shared points of Fig.~\ref{fig:FD2}).   } 
\label{fig:pencil}
\end{figure}

We assume the domain in the real space to be discretized into $N^3$ grid, which are divided evenly among $p$ processors  using pencil  decomposition, as shown in Fig.~\ref{fig:pencil}.  The processors themselves are divided equally along the $x$ and $y$ directions.  Hence, each processor has approximately $N/\sqrt{p}\times N/\sqrt{p} \times N$ points, as shown in Fig.~\ref{fig:pencil}.    In this figure, $p=p_x p_y$ with $p_x=4$ and $p_y=4$, and the processors indices varying from 0 to 15.

\begin{figure}
\centering
\includegraphics[scale=0.6]{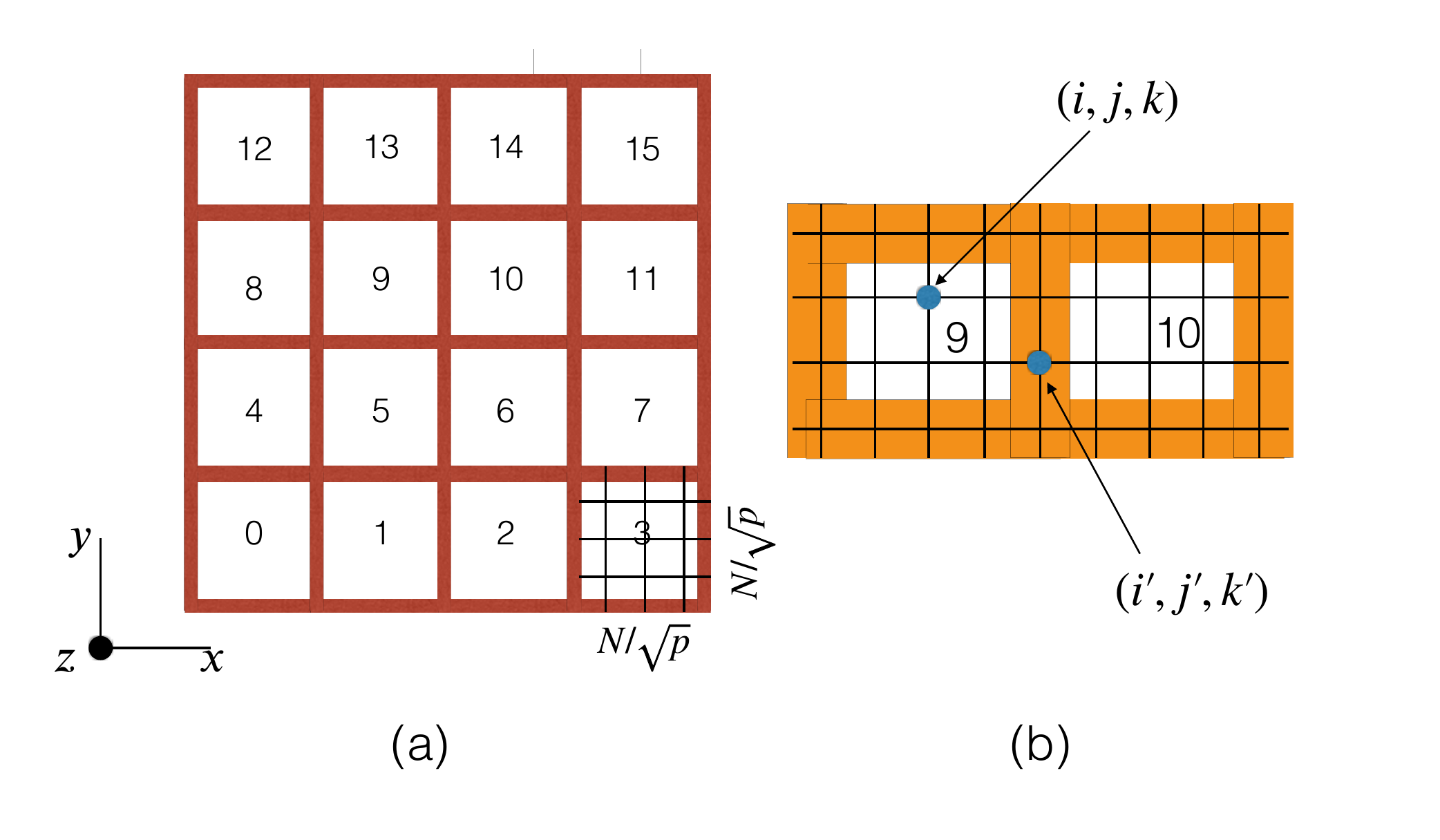}
\caption{ (a) The shared data among the processors (labelled as 0 to 15) is shown as  shaded regions.  (b) A zoomed view of the data decomposition among processors 9 and 10.  The grid point $(i,j,k)$ belongs  processor 9 alone, while the grid point $(i',j',k')$ belongs to processors 9 and 10. } 
\label{fig:FD2}
\end{figure}

In Fig.~\ref{fig:FD2} we illustrate how the data is shared among the processors.  Each processor shares data, such as the grid point $(i',j',k')$ of Fig.~\ref{fig:FD2}(b).   As a result of the shared data points, each processor contains slightly more data than $N/\sqrt{p}\times N/\sqrt{p} \times N$.   Sharing of data is important for derivative computation, as we describe below.

In the finite-difference scheme, the derivatives are approximate.  For example, a formula for $(\partial p/\partial x)_{(i,j,k)}$ in central difference scheme is
\be
\left( \frac{\partial p}{\partial x} \right)_{(i,j,k)} = 
\frac{p_{(i+1,j,k)} - p_{(i-1,j,k)}}{2 \Delta x}.
\ee
The above derivatives can be computed by a processor if both the points, $p_{(i+1,j,k)}$ and  $p_{(i-1,j,k)}$, are present in the processor.  However, the derivatives cannot be computed near the edges unless the data is shared among the neighbouring processors.   This is the reason why some data near the edges need to be shared among the processors.  The shared grid points are inside the shaded regions of Figs.~\ref{fig:FD2}(a,b); one of the shared points is  illustrated as $(i',j',k')$ in Fig.~\ref{fig:FD2}(b).  After the computation of the derivatives, the velocity field is time advanced.  For example, in Euler's scheme, Eq.~(\ref{eq:NS_u}) is time-advanced as
\be
{\bf u}_{(i,j,k)}(t+\Delta t) = {\bf u}_{(i,j,k)}(t) + (\Delta t) {\bf R}_{(i,j,k)}(t),
\ee
where the right-hand-side (RHS) term ${\bf R}$ is
\be
{\bf R}  =  -{\bf u} \cdot \nabla {\bf u}  - \nabla p +\nu \nabla^2 {\bf u}.
\ee

After each time step, each processor shares the updated field variables at the four interfaces  with four of its neighbouring processors. The amount of data to be shared are $O( N^2/\sqrt{p})$, where $O$ stands for ``of the order of".  It is easy to see that in the above scheme, the total amount of data to be transmitted is 
\be
D_\mathrm{FD} \approx 4\times 4 \times \frac{N^2}{\sqrt{p}} \times p \approx 16 N^2 \sqrt{p}.
\label{eq:data_comm_FD}
\ee
In the above formula, the factors $4 \times 4$ are for the 4 field variables ($u_x, u_y, u_z, p$) and for the 4 interfaces respectively.   

For the pencil decomposition shown in Fig.~\ref{fig:FD2},  Torus-2 (T2) is the most efficient interconnect because it facilitates communications among the neighbouring processors.  If sufficient number of ports (4  incoming and 4  outgoing) are available at compute nodes, direct connections among the neighbouring processors will minimise the communication time; this arrangement will be optimum for a small HPC cluster.  Also, it is best to implement a hybrid version---OpenMP for the cores within a node, and MPI for the communication across nodes; many finite difference codes have such arrangements.

A real implementation of a finite difference scheme involves many more steps.  For example, the pressure  for an incompressible flow is solved using a Poisson solver.  We refer the reader to Anderson~\cite{Anderson:book:CFD} and  Ferziger~\cite{Ferziger:book:CFD}  for these details. 

 In the next section, we briefly describe a pseudo-spectral method.

\section{Flow solvers based on pseudo-spectral scheme}
\label{sec:spectral}

\begin{figure}[htbp]
\begin{center}
\includegraphics[scale=0.5]{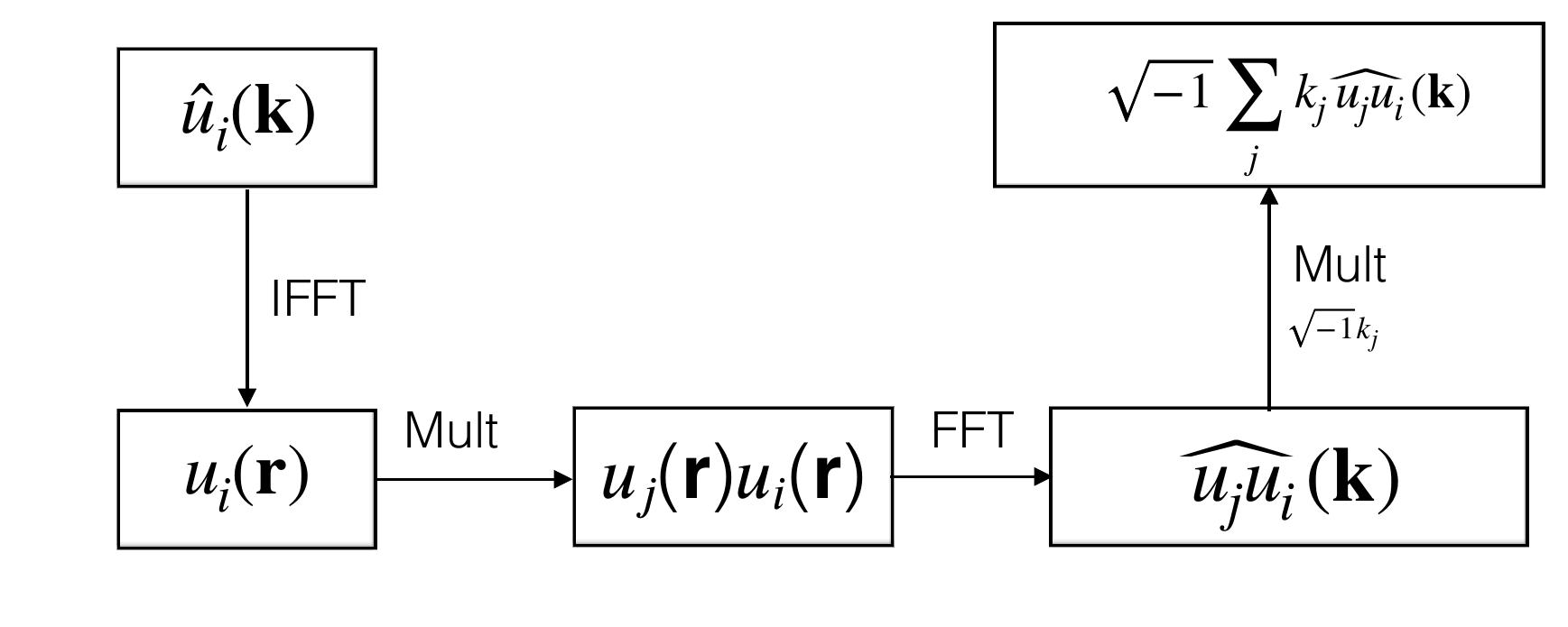}
\caption{The computation of the nonlinear term $\sqrt{-1} \sum_j  k_j \widehat{u_i u_j} ({\bf k})$ in a pseudo-spectral method.}
\label{fig:PS_method}
\end{center}
\end{figure}

In Fourier space,  Eqs.~(\ref{eq:NS_u},\ref{eq:div_u_eq_0ndim}) get transformed to
 \bea
  \frac{d}{d t}   \hat{u}_{i}(\mathbf{k},t) 
& = & - \sqrt{-1} k_{i} \hat{p} (\mathbf{k},t) -\hat{N}_{u,i}({\bf k})  -\nu k^2 \hat{u}_i({\bf k}),  \label{eq:ET:u_k_spectral}  \\
k_i  \hat{u}_{i}(\mathbf{k},t)  & = & 0,
\eea 
where ${\bf k}$ is the wavenumber, $\hat{f}$ is the Fourier transform of field $f$, and  $N_{u,i}$ is $i$th component of the nonlinear term:
\be
\hat{N}_{u,i} = \sqrt{-1} \sum_j k_j \widehat{u_j u_i} .
\ee
The nonlinear term is computed using Fast Fourier transform (FFT) to avoid convolution, whose computational complexity is $O(N^6)$ for a $N^3$ grid.  Comparatively, the computation complexity of a FFT is $O(N^3 \log N^3)$.  The steps involved in the computation process  are as follows (see Fig.~\ref{fig:PS_method})~\cite{Boyd:book:Spectral,Canuto:book:SpectralFluid,Verma:Pramana2013tarang,Chatterjee:JPDC2018}:
\begin{enumerate}
\item Compute ${\bf u(r)}$ from ${\bf \hat{u}(k)}$ using  inverse  FFT.
\item Compute ${u_{i}(\mathbf{r}) u_{j}(\mathbf{r})}$ in real space by multiplying the field variables at the space points.
\item Compute Fourier transform of $u_{i}(\mathbf{r})u_{j}(\mathbf{r})$ using forward FFT that yields $\widehat{(u_i u_j)}({\bf k})$.
\item Compute $\sqrt{-1}  \sum_j   k_{j}  \widehat{(u_i u_j)}({\bf k})$, which is the desired $\hat{N}_{u,i}({\bf k})$.
\end{enumerate}
Given $\hat{N}_{u,i}({\bf k})$, the pressure field is easily computed using
 \be 
 \hat{p} (\mathbf{k}) = \sqrt{-1} \frac{1}{k^2} \sum_j  k_j \hat{N}_{u,j}({\bf k}).
 \ee
The Fourier modes are time advanced using Euler or Runge Kutta schemes:
\be
{\bf u}(\mathbf{k},t+\Delta t) = {\bf u}(\mathbf{k},t) + (\Delta t) {\bf R}(\mathbf{k},t),
 \ee
 where
 \be
  {\bf R}(\mathbf{k},t) =- {\bf N}(\mathbf{k},t)  -\sqrt{-1} {\bf k} p(\mathbf{k},t)   - \nu k^2 {\bf u}(\mathbf{k},t) .
  \ee
  
 The most complex computation in the spectral method is the FFT, whose parallel implementation is described in the next section.  

\section{Parallel Computation of FFT}
\label{sec:FFTs}
For large $N$, we divide the data  among  $p$ processors using pencil decomposition, as shown in Fig.~\ref{fig:pencil}.   The forward and inverse FFT of the above data are defined as
\bea
\hat{f}(k_x, k_y, k_z) &= &  \sum_{x,y,z}  f(x,y,z)  \exp[-\sqrt{-1} (k_x x + k_y y+k_z z)], \\
f(x,y,z) &= &\sum_{k_x,k_y,k_z} \hat{f}(k_x, k_y, k_z) \exp[\sqrt{-1} (k_x x + k_y y+k_z z)].
\eea
These operations involve sums along  the three directions.  Note that an FFT computation involves all the data, hence it requires global communication among many processors.  This is contrary to the finite difference scheme that involves data transfers among the neighbouring processors.

\begin{figure*}[htbp]
\begin{center}
\includegraphics[scale = 0.7]{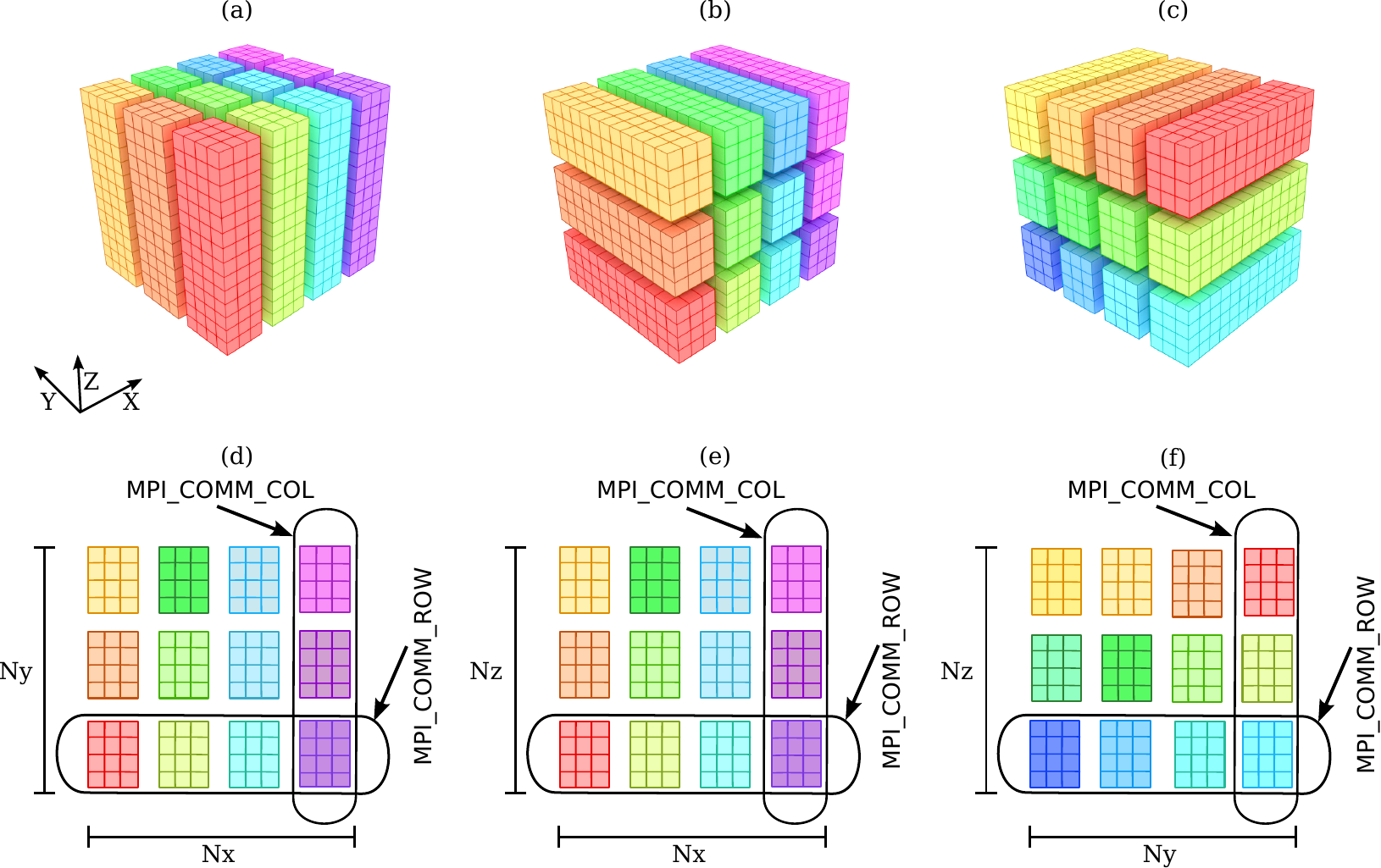}
\end{center}
\setlength{\abovecaptionskip}{0pt}
\caption{Illustration of data and operations involved during a Forward FFT transform:  (a) real space data, (b) intermediate configuration, (c) data in Fourier space.  (d, e, f) Division of cores into $p_\mathrm{row}$ and $p_\mathrm{col}$ with $p = p_\mathrm{row} \times p_\mathrm{col}$  as  in $XY$, $XZ$, and $YZ$ projections respectively.  Here $N_x = N_y =  N_z = 12$.  In the subfigures (a,d), $p_\mathrm{row} = 3$, $p_\mathrm{col} =  4$, thus each core contains $N_{x}/ p_\mathrm{col} \times N_{y}/ p_\mathrm{row}  \times N_{z} = 3\times 4 \times 12$ data points. From Chatterjee et al.~\cite{Chatterjee:JPDC2018}.  Reproduced with the permission from Springer Nature.}
\label{fig:pencil_combined}
\end{figure*}

In Fig.~\ref{fig:pencil_combined}, we illustrate the steps involved in forward transform (real space to Fourier space). In  this figure, the processors with same $X$ or $Y$ proc-coordinates form a set of communicators---\texttt{MPI\_COMM\_ROW} and \texttt{MPI\_COMM\_COL}.  Note that the division of processors in Fig.~\ref{fig:pencil_combined} is slightly different from that of Fig.~\ref{fig:pencil}.  Now the steps involved in a FFT operation are~\cite{Canuto:book:SpectralFluid,Chatterjee:JPDC2018}:
\begin{enumerate}
\item We perform one-dimensional (1D) forward FFT, \texttt{r2c} real-to-complex, along the Z-axis for each data column.
\item  We perform \texttt{MPI\_Alltoall} operation among the cores in  a \texttt{MPI\_COM\_COL} communicator to transform the data of Fig.~\ref{fig:pencil_combined}(a) to the intermediate configuration of Fig.~\ref{fig:pencil_combined}(b).  This process is repeated for all  \texttt{MPI\_COM\_COL} communicators.
\item After interprocess communication, we perform forward \texttt{c2c} (complex-to-complex) transform along the Y-axis for each pencil of the array. 
\item  We perform \texttt{MPI\_Alltoall} operation among the cores in a \texttt{MPI\_COM\_ROW} communicator to transform the data of Fig.~\ref{fig:pencil_combined}(b) to the Fourier configuration  of Fig.~\ref{fig:pencil_combined}(c). This process is repeated for all  \texttt{MPI\_COM\_ROW} communicators.  
\item  In the last step, we perform forward \texttt{c2c} transform along the X-axis for each  pencil [see Fig.~\ref{fig:pencil_combined}(c)].
\end{enumerate}
The \texttt{MPI\_Alltoall} communications are the most expensive operations  in the above.  Let us estimate the amount of data being communicated in an FFT operation.

Assuming equal division of processors along the X and Y directions, each processor has $N^3/p$ data.  In \texttt{MPI\_Alltoall}   communication within a  communicator, each processor sends and receives $\sqrt{p}-1$ packets of $N^3/(p\sqrt{p})$ data.  Hence, within each communicator, the amount of data exchanged is
\be
D_0 = \frac{N^3}{p\sqrt{p}} \frac{\sqrt{p} (\sqrt{p}-1)}{2}
\ee
Therefore, the total amount of data communicated across $\sqrt{p}$ communicator is
\be
D_\mathrm{PS} \approx \sqrt{p} D_0 \approx N^3
\label{eq:data_comm_FFT}
\ee
Hence, using Eqs.~(\ref{eq:data_comm_FD}, \ref{eq:data_comm_FFT}) we deduce that the ratio of data communicated in FFT and finite difference scheme is $O(N/\sqrt{p})$, which is large when $N \gg p$.  Note however that for FFT, the performance is much worse than the above ratio because \texttt{MPI\_Alltoall} communications across distant processors may require multi-hops in the interconnect. 

  FFT computations by various researchers show that data communication among the processors takes much longer than the computation time.  Here, we report the results of FFTK written by Chatterjee et al.~\cite{Chatterjee:JPDC2018}.    Chatterjee et al.  performed FFT on two parallel systems: Blue Gene/P (Shaheen I of KAUST), and Cray XC40 (Shaheen II of KAUST).   The Cray XC40 system has 6174 compute nodes each containing two Intel Haswell processors with 16 cores each.   In total, the system has a total of 197568 cores and 790 TB of memory. The compute nodes are connected via the Aries high-speed network, which is based on a dragonfly topology.  The Blue Gene/P supercomputer, an older system than Cray XC40,  had of 16 racks with each rack containing 1024 compute nodes having  32-bit 850-MHz quad-core PowerPC.   Hence the total number of cores in the system was 65536.  The Blue Gene/P nodes were connected via a three-dimensional Torus interconnect.  Note that Blue Gene/P system has been decommissioned.

\begin{figure}[htbp]
\begin{center}
\includegraphics[scale = 0.6]{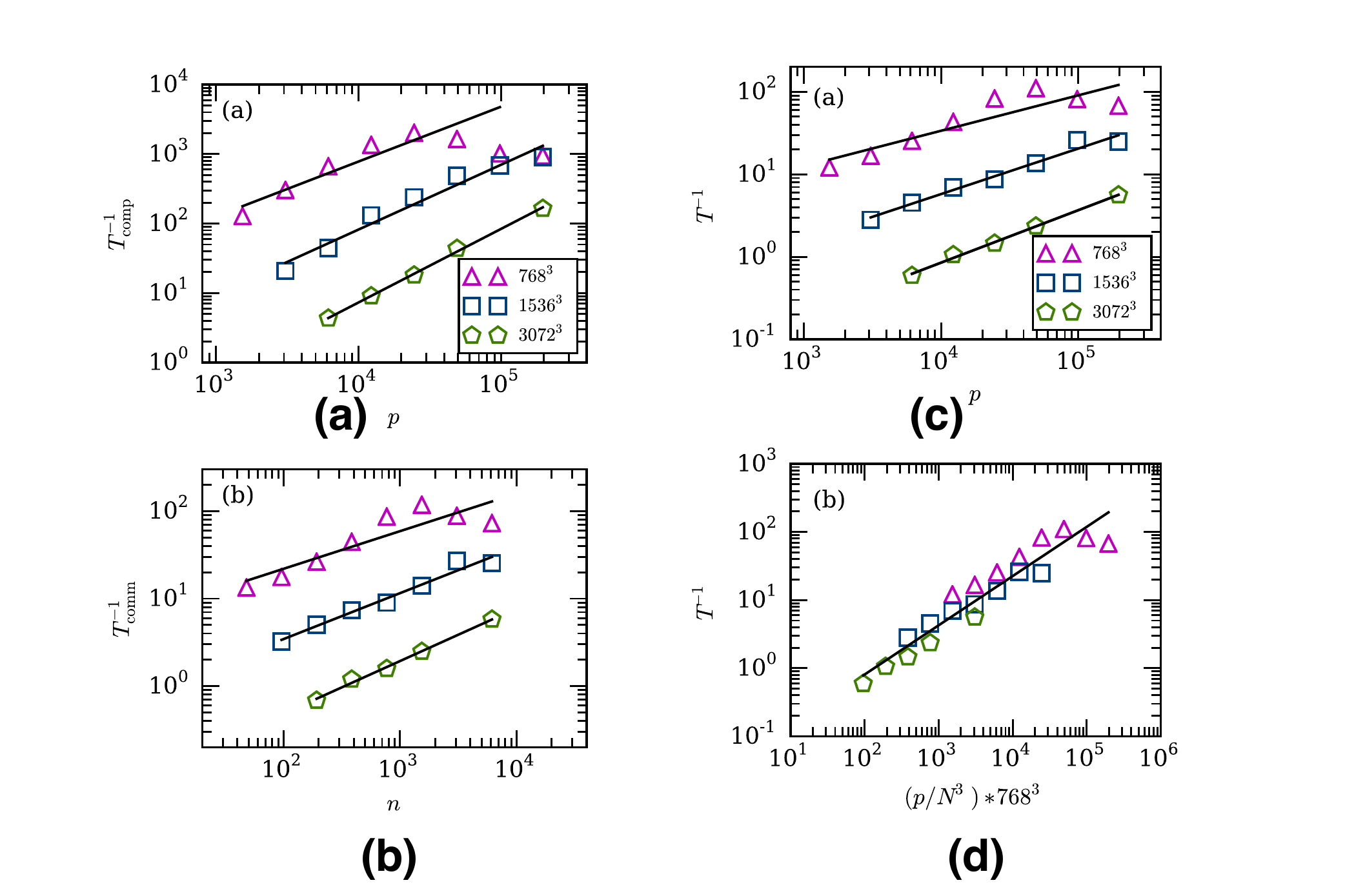}
\end{center}
\caption{Scalings of FFTK on Cray XC40: (a) Plots of  $T^{-1}_\mathrm{comp}$ vs.~$p$  (number of cores) for $768^3$, $1536^3$, and $3072^3$  grids.  (b) Plots of $T^{-1}_\mathrm{comm}$ vs.~$n$ (number of nodes) using the above convention.   (c) plots of  $T^{-1}$ vs.~$p$  for $768^3$, $1536^3$, and $3072^3$ grids.  (d) plots of $T^{-1}$ vs.~$p/N^3$ exhibits weak scaling with an exponent of $\gamma=0.72\pm0.03$.  Adopted from the figures of Chatterjee et al.~\cite{Chatterjee:JPDC2018}. Reprinted with permission from Springer Nature. }
\setlength{\abovecaptionskip}{0pt}
\label{fig:cray_fft}
\end{figure}

 For the runs on Cray XC40 for $768^3$, $1536^3$, and $3072^3$ grids, Chatterjee et al. computed the computation time, communication time, and total time taken for a pair FFT computation (forward and inverse).  They employed maximum of 196608 core, which are all the compute cores of Shaheen II.  The computation time decreases linearly with number of processors, i.e., $T^{-1}_\mathrm{comp} \sim p$.  Chatterjee et al.  characterised the communication time using an exponent $\gamma_2$: $T^{-1}_\mathrm{comm} \sim n^{\gamma_2}$, where $n$ is the number of nodes.  They found the exponent $\gamma_2$ for the three grids to be $0.43 \pm 0.09$, $0.52 \pm 0,04$, and $0.60 \pm 0.02$ respectively.  Since the communication time dominates the computation time, the exponents for the total time are close to $\gamma_2$.  The above scaling are illustrated in Fig.~\ref{fig:cray_fft}. Figure~\ref{fig:cray_fft}(c,d) shows respectively the strong and weak scaling for FFT.   

In addition to the above scaling, Chatterjee et al.~\cite{Chatterjee:JPDC2018}  estimated the efficiency of an FFT operation as the ratio of the effective FLOP (floating point operations) rating and the peak FLOP rating\footnote{The usual definition of efficiency, $T_\mathrm{serial}/(p T_\mathrm{parallel})$, is not suitable for large grids. This is because such large data cannot be accommodated within a single processor, hence, a sequential run for large grid is impossible. }.    The effective FLOP rating was estimated as the ratio of the total number of floating point operations and the total time taken.   For the three grids employed,  Chatterjee et al. reported the efficiencies to be 0.013, 0.015, and 0.018 respectively.  Such a low efficiency is due to the extreme data communication involved in FFT.  

Chatterjee et al.~\cite{Chatterjee:JPDC2018} carried out similar analysis on maximum of  65536 cores of Blue Gene/P, which is an older machine compared to Cray XC40.  They tested FFT scaling for $2048^3$, $4096^3$, and $8192^3$ grids using 1, 2, and 4 processors per node.  Surprisingly, Blue Gene/P yields better scaling--higher $\gamma_2$, and efficiency---than Cray XC40.   The peak efficiency of Blue Gene/P is 0.11, which is approximately 6 times the peak efficiency of Cray XC40.  See Chatterjee et al.~\cite{Chatterjee:JPDC2018}  for further details.  Note that each core of Cray XC40 is around 100 times faster than that of Blue Gene/P, however, the  interconnect speed of Cray XC40 has not improved in a similar proportion. Also, the Torus architecture of Blue Gene/P is better suited for \texttt{MPI\_Alltoall} communications than the Dragonfly architecture of Cray XC40. These are the prime factors for the lower efficiency of FFT on Cray XC40 compared to Blue Gene/P.      There could be other factors involving cache, memory access, etc. that needs to be examined carefully.  A lesson to be learnt from this exercise is that efficiency of a code depends on the speeds and architecture of processor, interconnect, memory, and cache---rather than processor alone. 

Let us contrast the above two results with those of Earth Simulator that operated in early 2000's.  Earth Simulator had 640 vector processors that were connected to each other via 640 x 640 crossbar switch and control units~\cite{Habata:NEC2003,Habata:PC2004}.   The crossbar interconnect offers an efficient implementation of  \texttt{MPI\_Alltoall}  communication with a single step.  This architecture led to remarkably efficient implementation of the spectral codes on Earth Simulator.  For example,  Earth Simulator achieved an efficiency of 64.9\% for  a global atmospheric circulation model, which is based the spectral method.  Note that on Earth Simulator,  the $N^3$ data  was divided into slabs because number of processors, 640, is much smaller than $N$, say 4096.   

Efficient spectral codes on Earth Simulator indicate a need for  specially-designed  and novel hardware for FFT. We may generalise the  efficient design of Earth Simulator to pencil decomposition, for which the optimum communication requires a fully-connected network within each communicator. Such schemes are  available neither in Torus nor in Dragonfly architecture.  We plan an approximate implementation of the above on Shaheen 2 in collaboration with KAUST's system team.  

FFTs take 60\% to  80\% of the total computer time in a spectral code. Hence, the efficiency of a spectral code is close to that of FFT. This is in addition to input/output of large data, which is typically implemented using parallel I/O, e.g. HDF5 library.

\section{Challenges in Implementation of CFD Codes in Exascale Systems}

In the last three sections we summarised the complexities of finite difference and spectral codes, as well as that of FFT.   From these examples, we can conclude that  in modern supercomputers, communication across interconnect is the  one of the leading  bottlenecks for the efficiency of CFD codes.  Among the two, a finite difference implementation is more efficient than a spectral one.  As described earlier, a finite difference code requires communication of much smaller dataset, that too among neighbouring processors. In comparison, the \texttt{MPI\_AlltoAll} communications in a spectral code requires transfer of much larger datasets among distant processors.  These communications may involve multi-hops within an interconnect.  These are  the primary reasons for the lower efficiency of a spectral code compared to a finite difference code.  Note however that a spectral code provides much better numerical accuracy than a finite-difference code.

Spectral simulations could be performed efficiently on Earth Simulator due to its interconnect with  a crossbar architecture.  Modern parallel computers  do not allow such possibilities because the crossbar architecture requires enormous connectivity ($n^2$ for $n$ nodes). Note however that modern compute nodes offer 8 processors, each having maximum of 32 cores.   A fully-connected network for limited number of such compute nodes will offer efficient implementation of FFT.   

Recent spectral simulations  employ a fraction of million processors. However, to best of our knowledge, the scaling studies on FFT have been performed up to maximum of 196608 processors~\cite{Chatterjee:JPDC2018}; in this study, the scaling curves tend to saturate near 196608 cores.    In addition,   FFT implementation on multi-GPUs  remains a major challenge due to communication issues.    Therefore, we may safely guess that present implementation of FFT will not scale beyond 1/2 million processors.

On the other hand, finite difference and finite volume schemes scale up to millions of cores.  For example,  Yang et al.~\cite{Yang:SC2016} ran a weather code on 10 million cores.  The efficiency for the explicit and implicit versions of their code are ~$\sim 100$\% and $\sim 52$\% respectively, which are much higher than those of spectral codes.  As described earlier,  higher efficiency for a finite difference code is due to its lower communication complexity. Hence, finite difference and finite volume schemes are better suited for exascale systems.  As described in Sec.~\ref{sec:FD}, Torus (T2) interconnect would provide an optimum data transfer for  finite difference codes.

Spectral elements~\cite{Deville:book} and Fourier continuation \cite{Albin:JCP2011} offer good promise for exascale computing.  These schemes provide flexibility of finite element/finite difference, as well as spectral accuracy.  In Fourier continuation, the real space domain (within a processor) is extended so as to make it periodic.  After this, accurate derivatives are computed by the respective processor using FFT, as in Sec.~\ref{sec:spectral}.  Since these derivatives are computed using partial data (within the processor), they are not as accurate as those of pseudo-spectral method.  But, there is an enormous  saving in communication cost.  In spectral element, the derivatives are computed using polynomials.  Thus, codes based on spectral elements and Fourier continuation could scale well in exascale systems. 

We also remark that shared memory architecture with hybrid implementation (OpenMP+MPI) provide interesting possibilities for efficient implementation of both spectral and finite difference codes.  Several exisisting codes~\cite{Rosenberg:PF2015}  employ such schemes.

In the next section we detail general computing issues in scientific applications.

\section{Challenges faced by an Application Scientist}
In this section, I will describe some of the difficulties faced by an application scientist in HPC.  The HPC technology is quite complex requiring expertise in software, hardware, and in application domain. On top of it,  it is very difficult to keep track of rapid development in the hardware and software technology, which are crucial for efficient implementation of application software.

The new processors  have large number of cores and bigger cache.  In addition, the modern interconnects are getting faster.  Exploitation of the above features  require  hybrid codes---OpenMP for the internal cores, and MPI for across nodes.   As described in this paper, appropriate network architecture and job scripts are required for efficient implementation of FFT and finite difference codes.   

Regarding software, large codes need to be structured and flexible (for frequent updates).  For the same, an application scientist needs to learn object-oriented programming.  Also, the features of parallel tools such as MPI and OpenMP are changing rapidly.  Porting the codes to multi-GPUs; and implementation of parallel I/O and version control are quite complex.  Lastly, some  computational algorithms (e.g. optimum cache usage)  are sometimes too complex for an application scientist. It is difficult to find physics and engineering  students who are skilled in these areas.  On top of it, there is pressure to deliver results in science and engineering domain.  Hence, one does not get sufficiently  long time for  development of efficient and robust codes.  

The above difficulties could be alleviated in a cross-disciplinary group having computational and application scientists.  Such groups are being formed these days, and we hope that they will become common in near future.

We  conclude in the next section.

\section{Conclusion and Discussion}

Futuristic exascale computers offer immense opportunities as well as challenges to application scientists.  In this paper we present computational challenges in computational fluid dynamics (CFD).  We present two generic schemes---finite difference and pseudo-spectral that involves FFT.   For FFT and pseudo-spectral codes,  inter-node communication is the biggest bottleneck in a generic HPC system.  Faster processors and relatively slower interconnects bring down the efficiency of a FFT code.  Experiences from Earth Simulator indicate that a fully connected network could yield high efficiency for FFT;  such network however would be very expensive.  Quantum Fourier Transform may offer an  alternative, but these discussions are beyond the scope of this paper~\cite{Yanofsky:book}

In contrast,  finite difference codes require communication of much smaller datasets, that too across neighbouring processors.  Hence, such codes are suitable for exascale HPC systems.  Yang et al.~\cite{Yang:SC2016} demonstrated how a finite-volume based weather code could be ported to 10 million cores.  For a better efficiency of such codes, it is important that the processor communicate among themselves in a single  hop, or in least possible hops.

A complex application involves many subsystems.  For example, weather codes have the following components: atmosphere, oceans, land, ice, etc.~\cite{Balaji:XRDS2013}.  For such codes, it is advisable to simulate the subsystems in different sets of processors, and then communicate the data among the subsystems.  Such software architecture would be robust, as well as less communication intensive.  Such codes too will be suitable for exascale systems.

Finally, there are design and documentation issues for  large codes.  All the above concerns need to be kept in mind while developing   large-scale CFD codes for exascale systems.

\section*{Acknowledgement}

The author thanks all the co-developers of FFTK, TARANG, and  finite difference code of our group.   Some of the key contributors to the codes are Anando Chatterjee, Rosan Samuel, Shaswat Bhattacharya, Ravi Samtaney, Fahad Anwer, Gaurav Gautam, Abhishek Kumar, Mani Chandra,  Akash Anand, Awanish Tiwari, and Soumyadeep Chatterjee.  In addition, author is grateful to  Akash Anand, Samar Aseeri, Rooh Khurram, Bilel Hadri, V. Balaji, and Preeti Malakar for discussion and ideas; and to  Ritu Arora, Venkatesh Shenoy, and Amitava Majumdar for organzing wonderful conference ``Software Challenges to Exascale Computing (SCEC)".  

Funding: This study was funded by research grants  INT/RUS/RSF/P-03 by the Department of Science and Technology India.  Our numerical simulations were performed on Cray XC40 (Shaheen II) and Blue Gene/P (Shaheen I) at KAUST supercomputing laboratory, Saudi Arabia, through project k1052.  

Conflict of Interest: The  author declares that he has no conflict of interest.
%
%

%
%
%

\end{document}